\documentclass[prb,twocolumn,superscriptaddress,showpacs,amsmath,amssymb]{revtex4}

\usepackage{graphicx}
\usepackage{latexsym}
\usepackage{amsmath}
\usepackage{amssymb}
\usepackage{amsfonts}
\usepackage{color}

\begin{document}

\title{Electronic Structure of Vortices Pinned by Columnar Defects.}

\author{A.~S.~Mel'nikov}
\author{A.~V.~Samokhvalov}
\author{M.~N.~Zubarev}
\affiliation{Institute for Physics of Microstructures, Russian
Academy of Sciences, 603950 Nizhny Novgorod, GSP-105, Russia}

\date{\today}
\begin{abstract}
The electronic structure of a vortex line trapped by an insulating
columnar defect in a type-II superconductor is analysed within the
Bogolubov-de Gennes theory.
For quasiparticle trajectories with small impact parameters defined with respect
to the vortex axis  the normal reflection of electrons and holes at the defect
surface results in the formation of an additional subgap spectral branch.
The increase in the impact parameter at this branch is accompanied by the decrease
of the excitation energy.
When the impact parameter exceeds the
radius of the defect this branch transforms into the Caroli--de Gennes--Matricon one.
As a result, the minigap in the quasiparticle spectrum increases with the increase in
the defect radius. The scenario of the spectrum transformation is generalized for the
case of arbitrary vorticity.
\end{abstract}


\maketitle

\section{Introduction}\label{Intro}

The study of  magnetic and transport properties of type-II superconductors in the presence of
artificial pinning centers is known to be an important direction in the physics of
vortex matter \cite{Blatter-RMP94}.
The artificial pinning provides a unique possibility to control the critical
parameters of superconducting materials which are important in various applications.
For instance, the critical current $j_c$ and the irreversibility field
$H_{irr}$ can be enhanced by the inclusion of normal particles and nanorods
\cite{Yang-Science96,Peurla-PRB07}, by introducing arrays of submicrometer holes
\cite{Hebard-IEEE77,Baert-PRL95}, and by proton \cite{Civale-PRL90} and heavy-ion
irradiation \cite{Civale-PRL91}. Pinning of flux lines appears to be especially
strong for the case of columnar defects elongated nearly parallel to the applied
magnetic field, when vortices can be pinned over their entire length.
These columnar defects are now widely used to trap vortices and to increase the
current carrying capacity of superconductors.

Within the London approximation
the interaction between a single vortex and an insulating cylindrical cavity
of radius $R \ll \lambda$, where $\lambda$ is the London penetration depth,
was considered in the pioneering paper
\cite{Mkrtchyan-JETP72} for bulk type-II superconductors.
For a multiquantum vortex it was shown \cite{Mkrtchyan-JETP72} that
 the maximum number of flux quanta which can be trapped by the cylindrical
  cavity is restricted by the value $R/2\xi$,
where $\xi$ is the superconducting coherence length.
The generalization of the results of Ref.~\cite{Mkrtchyan-JETP72}
for cylindrical cavities of  radii $R\gtrsim \lambda$ has been obtained in
Ref.~\cite{Nordborg-PRB00}.
An efficient image method appropriate for the analysis
of the vortex--defect interaction
in the limit of rather large $\lambda$ values
has been developed in \cite{Buzdin-PhysC96,Buzdin-PhysC98}.
The formation of superconducting nuclei with nonzero vorticities
near the columnar defects or in perforated films has been studied in Refs.
\cite{Buzdin-PRB93,Bezryadin-PLA94}.

Certainly the phenomenological approaches used in most of the works cited above can
not describe the electronic structure of the vortex states in the presence of small
cavities or colomnar defects of the radius smaller than the coherence length $\xi$.
This issue is closely related to the problem of microscopic nature of pinning
addressed previously in Ref.~\cite{thuneberg} for a particular case of point-like defects
with the scattering cross section much smaller than the $\xi^2$ value.
An appropriate modification of the quasiparticle spectra caused by a single impurity
atom placed in a vortex core has been studied in
\cite{Larkin-PRB98}.
The case of vortices trapped by normal metal cylindrical defects has been addressed in
\cite{Tanaka-JJAP95,Eschrig}.
The interest to microscopic calculations of electronic structure of the vortex states
is stimulated by low-temperature scanning tunneling microscopy (STM)
experiments which provide detailed spatially resolved excitation spectra
 \cite{Hess-PRL89,Hoogenboom-PRB00,Guillamon-PRL08}.
The modern STM techniques could provide us the information about the number and
 configuration of the spectral branches
crossing the Fermi level. Recent STM experiments on $\mathrm{Nb Se_2}$ single crystals
with a regular array of submicron $\mathrm{Au}$ antidots have provided images of both single quantum Abrikosov
vortices and multiquanta vortex states forming near normal antidots \cite{Karapetrov-PRL05}.

The goal of our paper is to analyse the transformation of the quasiparticle excitation spectra
which occurs in a vortex pinned by a columnar defect of finite radius
$R \lesssim \xi$.
 We focus on the
modification of the anomalous energy branches caused by normal reflection of
quasiparticles at the columnar defect boundary.
To elucidate the key points of the present work we start from the qualitative
discussion of the spectrum transformation scenario. Let us consider a vortex pinned at
an isolating cylinder of a radius $R$ (see Fig.1). The spectrum of quasiparticle
states can be analysed considering one--dimensional quantum mechanics of
electrons and holes along a set of linear quasiclassical trajectories.
Each trajectory is defined by the impact parameter $b$ and the trajectory orientation
angle (see Fig.1).
\begin{figure}[t!]
\includegraphics[width=0.3\textwidth]{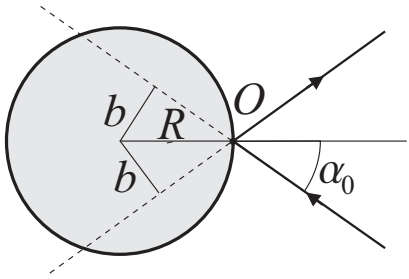}
\caption{Specular reflection of a quasiclassical
trajectory at the defect surface.}
\label{Fig:1}
\end{figure}
 For small impact parameters $b<R$ the trajectories
experience a normal reflection from the defect surface. Hereafter we assume this reflection to be
specular. Far from the reflection point $O$ the superconducting gap is homogeneous
($\Delta =\Delta_0$) and
the corresponding superconducting phase difference $\delta \varphi$ between the trajectory ends is
defined by the impact parameter $b$: $\delta \varphi = 2 \arcsin (|b|/R)$.
Neglecting the details of the inhomogeneous profile
of the order parameter inside the vortex core we can take
the gap function in  the form:
$\Delta (s) = \Delta_0 \exp (i \arcsin(|b|/R)\, {\rm sign}\,s)$, where
$s$ is the coordinate changing along the trajectory.
The one--dimensional quantum mechanical problem with such order parameter is
equivalent to the one describing a single mode Josephson constriction \cite{Beenakker}.
The subgap spectrum in this case is known to consist of  two energy branches:
$\varepsilon^\pm_{J} (b) = \pm \Delta_0\cos(\delta\varphi/2)=\pm\Delta_0\sqrt{1-b^2/R^2}$,
 which correspond to the opposite momenta
of quasiparticles propagating along the trajectory.
Thus, for small impact parameters the scattering of quasiparticles at the defect surface
 is expected to result in the formation of new energy branches which are splitted
from the continuum.
Taking the quasiclassical trajectories with large impact parameters $b>R$ one can see that these
trajectories are not perturbed by the scattering at the defect and, as a consequence,
the spectrum in this case should be described by the well-known Caroli-de Gennes-Matricon (CdGM)
expression \cite{Caroli-PL64}.
The crossover between two different regimes occurs in the region $b\sim \pm R$,
which should be certainly treated more accurately (see below).
One can expect that in this region the CdGM energy branch
$\varepsilon_{0} (b)$ is cut at the energies
$\varepsilon\sim \pm\varepsilon_{0} (R)$ and
transforms into the spectral branches $\varepsilon^\pm_{J}$
approaching $\pm\Delta_0$ with the further decrease in the $|b|$ value.
The resulting spectrum as a function of a continuous parameter $b$ does not cross the
Fermi level: there appears a minigap $\sim \varepsilon_{0} (R)$.
For $R\ll\xi$ this minigap can be approximately written as:
$\varepsilon_{0} (R)\simeq \Delta_0 R/\xi$.
The increase in the defect radius is accompanied by the minigap increase and for $R\gg\xi$
all the subgap states appear to be only weakly splitted from the $\pm\Delta_0$ value.

The paper is organized as follows. In Sec.~\ref{EqnSection} we briefly discuss the
basic equations used for the spectrum calculation. In Sec.~\ref{SingleQuant} we study
the quasiparticle spectrum transformation for a singly quantized vortex pinned at a
columnar defect. In Sec.~\ref{MultiQuant} we
generalize our analysis for the case of a multiquantum vortex
trapped by the defect.
We summarize our results in Sec.~\ref{sum}.


\section{Basic equations}\label{EqnSection}
Hereafter we consider a columnar defect as an insulating infinite cylinder
of the radius $R$. The magnetic field $\mathbf{B}$ is assumed to be parallel
to the cylinder axis $z$.
We assume the system to be homogeneous along the $z-$axis, thus,
the $k_z-$projection of the momentum is conserved.
The quantum mechanics of quasiparticle excitations in a superconductor
is governed by the two dimensional BdG equations for particlelike ($u$)
and holelike ($v$) parts of the two-component
quasiparticle wave functions
$\hat{\Psi}(\mathbf{r},z)=(u,\,v)\,\exp(i k_z z)$:
\begin{subequations}\label{eq:BdG}
\begin{eqnarray}
 -\frac{\hbar^2}{2m} \left( \nabla^2 + k_\perp^2 \right)\,u
      + \Delta(\mathbf{r})\, v  = \epsilon\, u\; \label{eq:BdGU} \\
\frac{\hbar^2}{2m} \left( \nabla^2 + k_\perp^2 \right)\,v
      + \Delta^*(\mathbf{r})\, u  = \epsilon\, v\;. \label{eq:BdGV}
\end{eqnarray}
\end{subequations}
Here $\nabla=\partial_x \mathbf{x}_0+\partial_y \mathbf{y}_0$,
$\mathbf{r}=(x,\;y)$ is a radius vector in the plane perpendicular to the cylinder axis, %
$\Delta(\mathbf{r})$ is the gap function, %
and $k_\perp^2=k_F^2-k_z^2$.

Following the procedure described in \cite{Melnikov-JETPL06,Kopnin-PRB07,Melnikov-PRB08}
we introduce the momentum representation:
\begin{equation}\label{eq:WFPR}
    \hat{\psi}(\mathbf{r}) =
        \left(u \atop v \right) =
        \frac{1}{(2\pi\hbar)^2}\int d^2\mathbf{p}\;%
            \mathrm{e}^{i\mathbf{p}\mathbf{r}/\hbar}\, \hat{\psi}(\mathbf{p})
\end{equation}
where $\mathbf{p} = \vert\mathbf{p}\vert\,%
    (\cos\theta_p\,, \sin\theta_p)= p\, \mathbf{p}_0$.
The unit vector $\mathbf{p}_0$ parametrized by
the angle $\theta_p$ defines the trajectory direction
in the ($x,\,y$) plane.
We assume that our solutions correspond to the momentum
absolute values $p$ close to the value $\hbar k_\perp$:
$p=\hbar k_\perp+q$ ($|q|\ll \hbar k_\perp$).
As a next step, we introduce a Fourier transformation:
\begin{equation}\label{eq:FTWFPR}
    \hat\psi(\mathbf{p})= \frac{1}{k_\perp}%
        \int\limits_{-\infty}^{+\infty} ds\, %
\mathrm{e}^{i(k_\perp-\vert\mathbf{p}\vert/\hbar)\,s}%
            \hat\psi(s,\theta_p)\,.
\end{equation}
Finally, the wave function in the real space
${\bf r}=r(\cos\theta,\sin\theta)$
is expressed from Eqs.(\ref{eq:WFPR}, \ref{eq:FTWFPR})
in the following way
(see Ref.\cite{Kopnin-PRB07}):
\begin{equation}\label{eq:WFCS}
    \hat{\psi}(r,\theta) %
        =\int\limits_{0}^{2\pi} %
            \mathrm{e}^{i k_\perp r cos(\theta_p-\theta)} %
            \hat\psi(r \cos(\theta_p-\theta),\theta_p) %
            \frac{d\theta_p}{2\pi}\,,
\end{equation}
where ($r$, $\theta$, $z$) is a cylindrical coordinate
system.
The boundary condition at the surface of the insulating
cylinder requires
\begin{eqnarray}
    &&\hat{\psi}(R,\theta) ={\left(u \atop v \right)}_{r=R}= \label{eq:WFBC}\\
    &&\quad\frac{ 1}{2\pi}
        \int\limits_{0}^{2\pi}d\theta_p
        \mathrm{e}^{i k_\perp R cos(\theta_p-\theta)}
        \hat\psi(R \cos(\theta_p-\theta),\theta_p) = 0\,. \nonumber
\end{eqnarray}

To obtain the Andreev equations along the trajectories
we look for a solution in the eikonal approximation
$$
\hat{\psi}(s,\theta_p) = \mathrm{e}^{i S(\theta_p)} \hat{g}(s,\theta_p)
$$
assuming $\hat{g}$ to be a slowly varying function of $\theta_p$.
Quasiparticles propagating along the classical trajectories
parallel to $\mathbf{k}_\perp = k_\perp(\cos\theta_p, \sin\theta_p)$
are characterized  by the angular momentum
$\mu=-k_\perp b$, where
\begin{equation}\label{eq:IP}
    b = -\frac{1}{k_\perp}\frac{\partial S}{\partial \theta_p}
\end{equation}
is the trajectory impact parameter.
Assuming the vortex axis to coincide with the cylinder axis we obtain
the axially-symmetric problem with the conserved angular momentum $\mu$.

Finally, the quasiclassical equations for the
envelope $\hat{g}(s,\theta_p)$ read:
\begin{eqnarray}
    -i \hbar V_\perp \hat{\sigma}_z %
               \frac{\partial\hat{g}}{\partial s}%
        &+&\hat{\sigma}_x \mathrm{Re}\Delta(\mathbf{r})\,\hat{g}\quad   \label{eq:QCAE} \\%
        &-&\hat{\sigma}_y \mathrm{Im}\Delta(\mathbf{r})\,\hat{g}%
=\left(\epsilon+\frac{\hbar\,\omega_H}{2}\,\mu\right)\hat{g}\,,\nonumber
\end{eqnarray}
where $\hat{\sigma}_i$ 
are the Pauli matrices,
$m V_\perp = \hbar k_\perp$,
$\omega_H=\vert e \vert H / m c$ is the cyclotron frequency, and
\begin{eqnarray}
x = s\cos\theta_p- b\sin\theta_p\,,
\quad
y = s\sin\theta_p + b\cos\theta_p\,,       \nonumber \\
x \pm i y = (s \pm i b)\,%
    \mathrm{e}^{\pm i \theta_p}\,.\qquad\qquad  \nonumber
\end{eqnarray}
The term proportional to $\omega_H$ can be included
 to the energy as an additive constant (see also \cite{hansen}):
$$
\varepsilon=\epsilon+\frac{\hbar\,\omega_H}{2}\,\mu.
$$
Our further analysis of quasiparticle excitations
is based on the Andreev equations (\ref{eq:QCAE})
which must be supplemented by the boundary condition
(\ref{eq:WFBC}).


\section{Singly quantized vortex pinned
by a columnar defect}\label{SingleQuant}

We now proceed with the
analysis of the subgap spectrum for a
singly quantized vortex trapped
by the columnar defect of the radius $R$.
The order parameter $\Delta(x,y)$ takes the form
\begin{equation}\label{eq:OP}
\Delta = \Delta_0\, \delta_v(r)\,\mathrm{e}^{i\theta}\,,%
    \quad r=\sqrt{x^2+y^2}\ge R\,.
\end{equation}
Here $\delta_v(r)$ is a normalized order parameter magnitude for
a vortex centered at $r=0$, such that $\delta_v(r) = 1$ for $r\to\infty$.
In ($s$, $\theta_p$) variables one obtains for $r=\sqrt{s^2+b^2} \ge R$:
\begin{equation}\label{eq:OPST}
    \Delta = D_b(s)\,\mathrm{e}^{i\theta_p}\,,\quad
             D_b(s)=\Delta_0\,\frac{\delta_v(\sqrt{s^2+b^2})}
                                   {\sqrt{s^2+b^2}}(s+i b)\,.
\end{equation}
The cylindrical symmetry of our system allows to separate
the $\theta_p-$dependence of
the function $\hat{g}$:
\begin{equation}\label{eq:QCE0-F}
\hat{g}(s,\theta_p) = %
     \mathrm{e}^{i\,\hat{\sigma}_z \theta_p /\, 2}\hat{f}(s) \ .
\end{equation}
The total wave function $\hat{\psi}(s,\theta_p)$
should be single valued and, thus, the angular momentum
$\mu$ is half an odd integer.
The quasiclassical equations (\ref{eq:QCAE}) take the form
\begin{equation}\label{eq:QCE0}
    -i \hbar V_\perp \hat{\sigma}_z\,\partial_s\hat{f}
              + \hat{\Delta}_b(s) \hat{f} = \varepsilon \hat{f}\,,
\end{equation}
where
\begin{equation}\label{eq:QCE0-D}
    \hat{\Delta}_b(s) = \hat{\sigma}_x\,\mathrm{Re}D_b(s) %
                       -\hat{\sigma}_y\,\mathrm{Im}D_b(s)
\end{equation}
is the gap operator.
Changing the sign of the coordinate $s$ one can observe
 a useful symmetry property of the solution of Eq.(\ref{eq:QCE0}):
\begin{equation}\label{eq:QCE0-SYM}
\hat f (-s) = C\,\hat{\sigma}_y \hat f (s) \ ,
\end{equation}
where $C$ is an arbitrary constant.

\subsection{Boundary condition.}\label{BounCond}
As a next step we rewrite the boundary condition (\ref{eq:WFBC})
for wave functions $\hat{f}(s)$ defined at the trajectories.
Replacing $\theta_p$ by $\alpha = \theta_p-\theta$ and shifting the limits of
integration in Eq.(\ref{eq:WFBC}) we find:
\begin{equation}\label{eq:WFBC1}
    \int\limits_{0}^{2\pi}d\alpha\,
    \mathrm{e}^{i k_\perp R \cos\alpha + i\mu\alpha}
    \left[\mathrm{e}^{i\hat\sigma_z\alpha/2}%
    \hat{f}(R \cos\alpha)\right] = 0\, .
\end{equation}
Assuming $k_\perp R \gg 1$ and the
function $\mathrm{e}^{i\hat\sigma_z\alpha/2}\hat{f}(s)$
to vary slowly at the atomic length scale
we evaluate the above integral using the stationary phase method.
For a given value of angular momentum $\mu$
the stationary phase points are given
by the condition: $\sin\alpha_{1,2}=\mu/k_\perp R=-b/R$.
One can see that for $|b| > R$ the stationary phase points disappear
and, as a result, the integral (\ref{eq:WFBC1}) is always vanishingly
small. In this case the boundary condition
at the cylinder surface does not impose any restrictions
on the wave function $\hat f$ defined at the trajectories.
In the opposite limit  $|b| < R$
one can find two stationary angles $\alpha_1=\alpha_0\equiv-\arcsin(b/R)$
and $\alpha_2=\pi-\alpha_0$ which
are in fact the orientation angles for an incident and specularly reflected
trajectories shown in Fig.~\ref{Fig:1}.
Summing over two contributions we can rewrite the boundary condition
(\ref{eq:WFBC1})
as follows:
\begin{equation}\label{eq:WFBC2}
    \mathrm{e}^{ i\hat{\varphi}_1}\hat f(s_0)=
    \mathrm{e}^{-i\hat{\varphi}_1}\hat f(-s_0)\, ,
\end{equation}
where $s_0=\sqrt{R^2-b^2}$ , $2\beta_0 = \alpha_0-\pi/2$  and
$$
\hat{\varphi}_1 = k_\perp s_0 +
    (2\mu+\hat\sigma_z)\beta_0 - 3\pi/4\, .
$$

\subsection{Solution for large impact parameters $|b|>R$.}\label{LargeB}

In this case the quasiparticle states at the trajectories are not affected by the
 the normal scattering at the columnar defect boundary and the
behavior of an anomalous energy branch is described by
the standard CdGM solution for a single Abrikosov vortex.
For the sake of completeness we give below the expressions for this spectrum and
the corresponding wave functions.

Let us follow the derivation in Ref.~\cite{Volovik-JETPL93}
and consider the imaginary part of the gap operator (\ref{eq:QCE0-D})
as a perturbation.
Neglecting this term in Eq.~(\ref{eq:QCE0}) we find:
\begin{equation}\label{eq:QCE1Z}
    -i \hbar V_\perp \hat{\sigma}_z \partial_s\hat{f}_0
              + \hat{\sigma}_x\,\mathrm{Re}D_b(s)\,\hat{f}_0
    = \varepsilon\hat{f}_0\,.
\end{equation}
The above equation has a zero eigenvalue $\varepsilon=0$
with the following expression for the
corresponding normalized eigenfunction $\hat{f}_0$:
\begin{equation}\label{eq:QCE1Z-F}
    \hat{f}_0 =\sqrt{\frac{1}{2I}} %
                    {1\choose -i}\,
                    \mathrm{e}^{-K_0(s)}\,,
\end{equation}
where
\begin{equation}\label{eq:QCE1Z-K}
K_0(s) = \frac{1}{\hbar V_\perp}\int\limits_0^{s} dt\,%
                                \mathrm{Re}\,D_b(t)\,, \quad
I_0= \int\limits_{-\infty}^{+\infty} ds\,\mathrm{e}^{-2K_0(s)}\,.
\end{equation}
The first order perturbation theory gives us the CdGM excitation
 spectrum $\varepsilon_0(b)$ for $\vert b \vert > R$:
\begin{equation}\label{eq:CdGM}
    \varepsilon_0(b) =
        \frac{ b\,\Delta_0 }{I_0} \int\limits_{-\infty}^{+\infty}ds\,
        \frac{ \delta_v \left(\sqrt{s^{2}+b^{2}}\right)}
             {\sqrt{s^{2}+b^{2}}}\,
             \mathrm{e}^{-2K_0(s)}\,.
\end{equation}

\subsection{Solution for small impact parameters $|b|<R$.}\label{SmallB}

In this case the specular reflection at the cylinder surface
changes the trajectory direction and strongly modifies the spectrum.
The boundary condition at the surface $r=R$
is determined by the equation (\ref{eq:WFBC2}).
Let us introduce the function
\begin{equation}\label{eq:QCE2-F}
\hat{F}(s)=\left\{
        {\mathrm{e}^{+i\hat{\varphi}_1}\hat{f}(s + s_0)\,,\quad s > 0}
        \atop%
        {\mathrm{e}^{-i\hat{\varphi}_1}\hat{f}(s - s_0)\,,\quad s < 0}
           \right.  \ ,
\end{equation}
which is defined at the full  $s$ axis and appears to be
continuous at $s=0$: $\hat{F}(-0)=\hat{F}(+0)$.
The equation for $\hat{F}$ reads:
\begin{eqnarray}
    -i\hbar V_\perp \hat{\sigma}_z\,\partial_s \hat{F}%
            &+&\hat{\sigma}_x\,\mathrm{Re}\,G(s)\,\hat{F} \label{eq:QCE2}\\
            &-&\hat{\sigma}_y\,\mathrm{Im}\,G(s)\,\hat{F}
    = \varepsilon\hat{F} \ ,                   \nonumber
\end{eqnarray}
where
\begin{eqnarray}
   &&G(s) =  -\Delta_0\,\frac{\delta_v(\sqrt{(|s|+ s_0)^2+b^2})}%
                                  {\sqrt{(|s|+ s_0)^2+b^2}} \label{eq:QCE2-G}\\
                             &&\qquad\qquad\times
                               \left[
                                      s b / R + i \left(R +|s| \sqrt{1-b^2 / R^2} \right)
                               \right]\,. \nonumber
\end{eqnarray}
Taking the limit of large $|s|$ we find:
\begin{equation}
    G(s) =  -i \Delta_0\, \mathrm{e}^{\,i\,\alpha_0\,| s | / s } \,.
\end{equation}
One can see that in agreement with the qualitative arguments given in the Introduction
the phase difference between the opposite ends of the trajectory equals to
$\delta\varphi = -2\alpha_0 = 2 \arcsin( b / R )$.
Provided we neglect the inhomogeneity of the order parameter phase inside the core
we find the resulting expression for the spectrum:
$\varepsilon = \pm\Delta_0\sqrt{1-b^2/ R^2} $.
Certainly, such simplification does not allow us to study the crossover to
the CdGM branch which occurs at $b\sim \pm R$.
To develop an analytical description
of this crossover we choose to apply the method used above to derive standard CdGM
expressions and based on the perturbation theory with respect to the imaginary part of
the gap function. One can expect this method to be most adequate for the
crossover region of $b\sim R$. As for the limit $b< R$ we shall check the validity
of this method using the comparison with our direct numerical analysis of Eq.(\ref{eq:QCE2}).

Neglecting the maginary part of $G$ we find an exact solution of the equation
(\ref{eq:QCE2}) corresponding to zero energy $\varepsilon=0$:
\begin{equation}
    \hat{F}_{0}(s) = \sqrt{\frac{1}{2 I}} {1\choose i\,\chi}\, %
                   \mathrm{e}^{-K(s)}\,, \label{eq:QCE2Z-F}
\end{equation}
where
\begin{equation} \label{eq:QCE2Z-KI}
    K(s) = \frac{\chi}{\hbar V_\perp}\int\limits_0^{s} dt\, %
                   \mathrm{Re}\,G(t)\,,\quad
    I = \int\limits_{-\infty}^{+\infty} ds\, e^{-2K(s)}\,,
\end{equation}
%
and $\chi = \mathrm{sign}\,b$.
The solutions (\ref{eq:QCE2Z-F}), (\ref{eq:QCE2Z-KI})
appear to decay both at negative and positive $s$ and, thus, we
get a localized wave function describing a bound state.
Using this localized solution as a zero--order approximation for
the wave function the spectrum can be found within the first order
perturbation theory.
Note, that our perturbation procedure fails for $\vert\, b\, \vert \to 0$
because of the increase in the localization radius of the wave function
(\ref{eq:QCE2Z-F}).

The first--order approximate solution of the
quasiclassical equations (\ref{eq:QCE2}) takes the form:
\begin{eqnarray}
    \hat{F}(s) = A\left(1\atop i\chi\right)%
                                \mathrm{e}^{-K(s)}
                     + B(s)\left(1\atop -i\chi\right)%
                                \mathrm{e}^{K(s)}, \label{eq:QCE2-FA}
\end{eqnarray}
where
\begin{equation}
    B(s) = \frac{i A}{\hbar V_\perp}%
                    \int\limits_{-\infty}^{s} dt\,
                    \left[ %
                        \varepsilon - \chi\,\mathrm{Im}\,G(t)
                    \right]\,\mathrm{e}^{-2 K(t)}\,. \label{eq:QCE2Z-B}
\end{equation}
To avoid the wave function divergence we should put
$$
    \int\limits_{-\infty}^{\infty} dt\,%
        \left[ \varepsilon - \chi\,\mathrm{Im}\,G(t)
        \right]\,\mathrm{e}^{-2 K(t)} = 0\,.
$$
This condition gives us the excitation spectrum $\varepsilon_s$
as a function of the impact parameter $b$ for $\vert b \vert < R$ :
\begin{eqnarray}
    \varepsilon_s(b) &=&
        \frac{ \chi\, \Delta_0 }{I} %
            \int\limits_{-\infty}^{+\infty} ds\,
                \frac{\delta_v(\sqrt{(| s |+s_0)^2+b^2})}%
                                  {\sqrt{(| s |+s_0)^2+b^2}} \label{eq:QCE2-E}\\
                             &&\qquad\qquad\times
                              \left( R + | s | \sqrt{1-b^2 / R^2} \right)
                                \mathrm{e}^{-2 K(s)}\,. \nonumber
\end{eqnarray}
It is evident that $\varepsilon_s(R) = \varepsilon_0(R)$ and, thus, the expressions (\ref{eq:CdGM}) and (\ref{eq:QCE2-E}) describe the spectrum $\varepsilon(b)$ for an arbitrary
impact parameter $b$ :
\begin{equation}\label{eq:QCE2-EF}
    \varepsilon(b)=\left\{
        {\varepsilon_s(b)\,,\quad \vert\, b\, \vert \le R}
        \atop%
        {\varepsilon_0(b)\,,\quad \vert\, b\, \vert > R}
           \right.\,.
\end{equation}
The discontinuity of the derivative $d\varepsilon/db$
at $\vert\, b\, \vert = R$ appears because of the breakdown of the above quasiclassical
description for the rectilinear trajectories touching
the surface of the defect.

%
\begin{figure}[b!]
\includegraphics[width=0.45\textwidth]{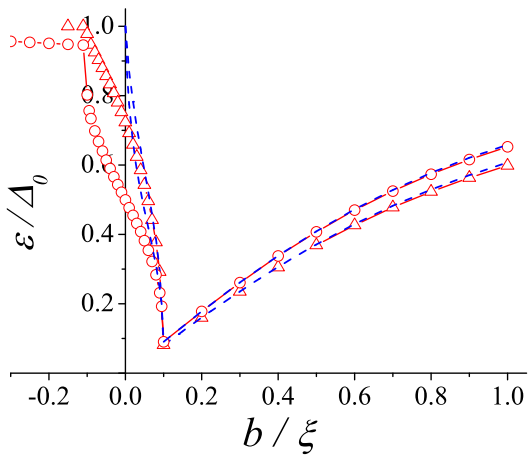}
\caption{(Color online) The spectral branches as  functions of the impact parameter $b$,
obtained from numerical solution of the
eigenvalue problem (\ref{eq:QCE0}),(\ref{eq:WFBC3}), are shown by red
triangles and circles for $k_z = 0$ and $k_z = 0.9 k_F$ , respectively.
The spectral branches calculated using Eq.~(\ref{eq:QCE2-EF})
are shown by blue dash lines. Here we put $R = 0.1\xi$.}
\label{Fig:2}
\end{figure}
%
\begin{figure}[t!]
\includegraphics[width=0.50\textwidth]{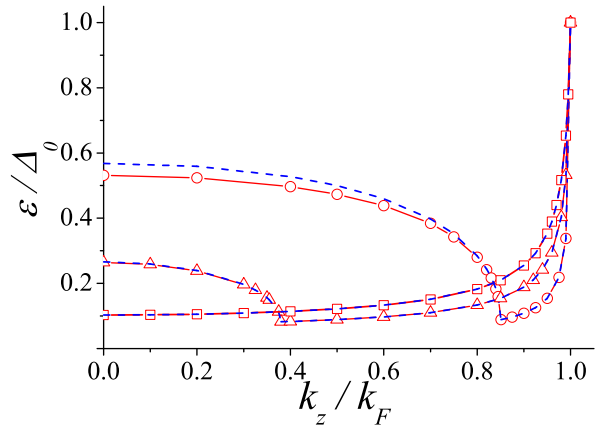}
\caption{(Color online) The quasiparticle spectra obtained from the numerical solution of the
eigenvalue problem (\ref{eq:QCE0}), (\ref{eq:WFBC3}) are shown
by red circles ($\mu=-10.5$), triangles ($\mu=-18.5$), and squares ($\mu=-25.5$).
The spectral branches calculated using Eq.~(\ref{eq:QCE2-EF})
are shown by blue dash lines. Here we put $R = 0.1\xi$, $k_F \xi = 200$.}
\label{Fig:3}
\end{figure}
%
In figures~2 and 3 we compare the typical plots of
quasiparticle spectra obtained
analytically, i.e., using Eq.(\ref{eq:QCE2-EF}), and
numerically.
The qualitative behavior of the spectrum is weakly sensitive to the
concrete gap profile inside the core and, thus, we choose a simple model profile:
\begin{equation}\label{eq:CORE}
    \delta_v(r)= r / \sqrt{r^2+\xi_v^2} \ ,
\end{equation}
where the core size $\xi_v$ equals to the coherence length
$\xi = \hbar V_F / \Delta_0$.
We plot here only the spectrum for positive energies and $k_z$ momenta
because the eigenvalues for $\varepsilon < 0$ and $k_z < 0$
can be found using the spectrum symmetry properties:
$\varepsilon(-b,k_z) = -\varepsilon(b,k_z)$
and $\varepsilon(b,-k_z) = \varepsilon(b,k_z)$.
To find the spectral branch $\varepsilon(b,k_z)$ numerically
we solve quasiclassical equations (\ref{eq:QCE0})
for $s \ge s_0$ requiring the decay of the wave function $\hat{f}$ at $s \to \infty$.
An appropriate boundary condition for electron $f_u$ and hole $f_v$
components of the wave function $\hat{f}=(f_u,f_v)$
at $s=0$ can be found from Eq.(\ref{eq:WFBC2}) and the symmetry property
(\ref{eq:QCE0-SYM}):
\begin{equation}\label{eq:WFBC3}
    f_v(s_0) = e^{i \alpha_0} f_u(s_0).
\end{equation}
For $\vert b \vert > R$ we put $s_0 = 0$, and
the boundary condition (\ref{eq:WFBC3}) takes
the form $f_v(0)= -i\,\chi\,f_u(0)$.

Comparing the spectrum (\ref{eq:QCE2-EF}) with the branches
obtained from the direct numerical analysis
of the eigenvalue problem (\ref{eq:QCE0}), (\ref{eq:WFBC3}) one can
see that the  perturbation method provides a reasonable
description of the energy spectrum behavior
in a wide range of the impact parameters.
As one would expect, the perturbation procedure fails
for small impact parameters $\vert\, b\, \vert \ll R$.
Contrary to the CdGM case the  spectrum branch (\ref{eq:QCE2-EF})
does not cross the Fermi level, and the minigap in the quasiparticle
spectrum $\Delta_{min}= \varepsilon(R)$ grows with the increase in the
cylinder radius $R$ (see Fig.~2).
Existence of the minigap in the spectrum of quasiparticles
should result in peculiarities of the density of states (DOS) and can
be probed by the STM measurements.
For $\vert\, \mu\, \vert <  k_F R$ the spectrum $\varepsilon(k_z)$
has a minimum (see Fig.~3), therefore we can expect the appearance
of a van Hove singularity in the energy dependence of the DOS.


\section{Quasiparticle spectrum of a multiquantum vortex}
\label{MultiQuant}

%
\begin{figure*}[t!]
\includegraphics[width=0.47\textwidth]{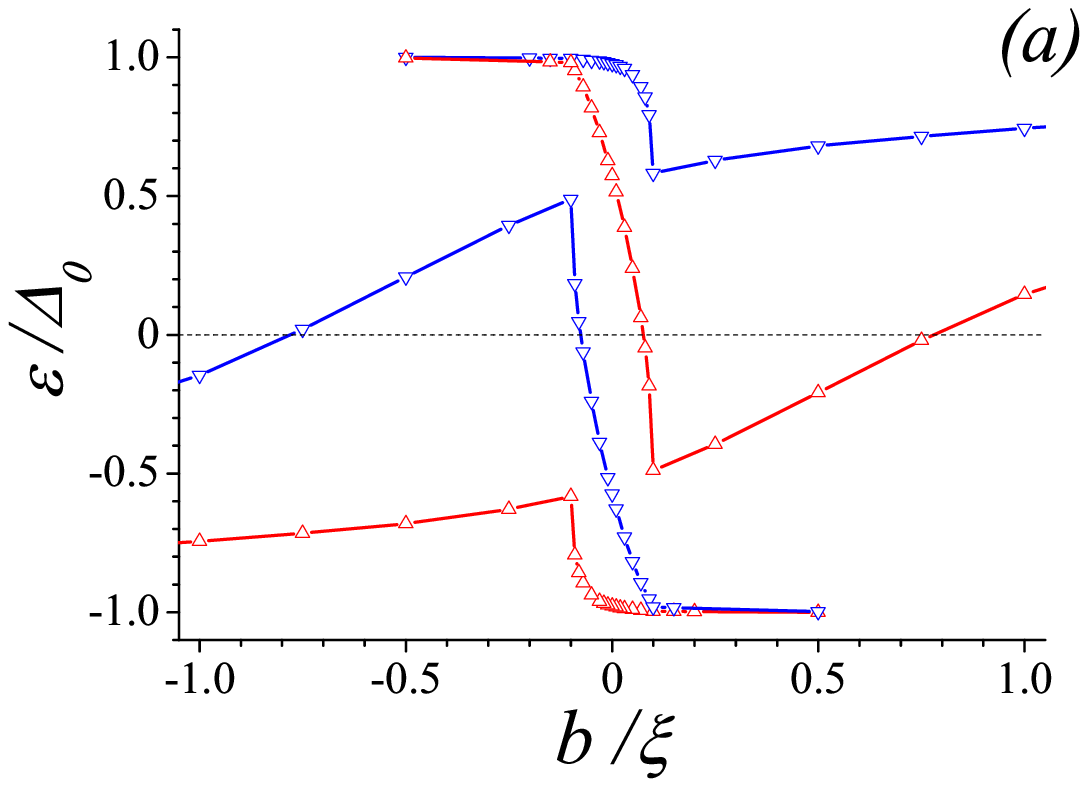}
\includegraphics[width=0.47\textwidth]{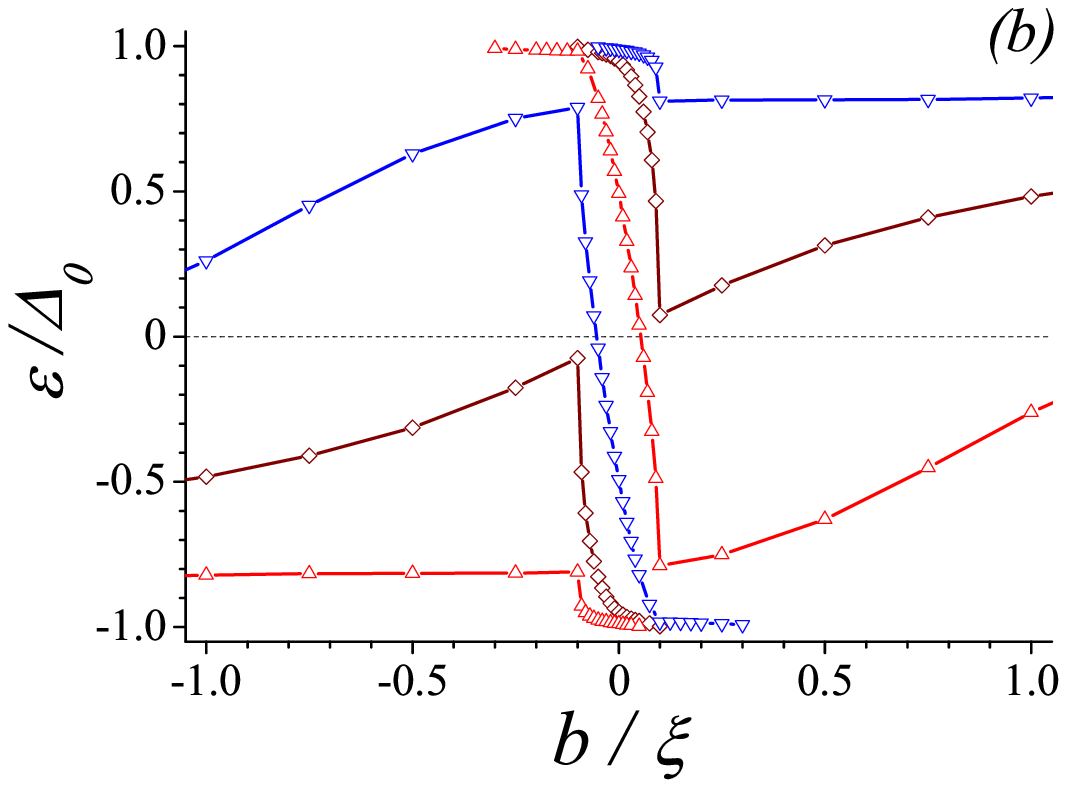}
\caption{(Color online) The  spectral branches as functions of the impact parameter $b$,
obtained from the numerical solution of the
eigenvalue problem (\ref{eq:QCE0}), (\ref{eq:QCE0-DM})
for $M=2$ (a) and $M=3$ (b) ($R = 0.1\xi$) .}
\label{Fig:4}
\end{figure*}
%
In this section we generalize the above analysis
for the case of a multiquantum vortex pinned by the columnar defect of the radius $R$.
The multiquantum vortices can be trapped at columnar defects either for a rather large defect radius
or for the mixed state in mesoscopic samples
\cite{Karapetrov-PRL05,Bezryadin-PRB96,Silhanek-PRB04,Grigorieva-PRL07}.
In the absence of defects the spectrum of a multiquantum vortex with the vorticity $M$
is known to consist of $M$ anomalous energy branches \cite{Volovik-JETPL93}.
The behavior of these  branches has been previously investigated both numerically and analytically
\cite{Tanaka-JJAP95,MQ-AnBranchSimul,Eschrig,Melnikov-Nature02,Melnikov-PRB08}.
Here we restrict ourselves by the numerical solution of the eigenvalue problem
(\ref{eq:QCE0}) assuming that the order parameter $\Delta_M(r)$ takes the form
\begin{equation}\label{eq:OPM}
    \Delta_M(r) = \Delta_0\, [\,\delta_v(r)\,]^M %
             \mathrm{e}^{i M \theta}\,,%
    \quad r \ge R\,,
\end{equation}
where the function $\delta_v(r)$ is determined by the expression (\ref{eq:CORE}).
In ($s$, $\theta_p$) variables one obtains for $s \ge s_0$:
\begin{eqnarray}
    &&\Delta_M = D_M(s)\,\mathrm{e}^{i M \theta_p}\,, \label{eq:OPSTM1}\\
    &&D_M(s)=\Delta_0\,\left[ \frac{\delta_v( \sqrt{s^2+b^2} )} %
                                  {\sqrt{s^2+b^2}} \right]^M %
                                  {(s+i b)}^M\,. \label{eq:OPSTM2}
\end{eqnarray}
Using the transformation
\begin{equation}\label{eq:QCE0-FM}
\hat{g}(s,\theta_p) = %
     \mathrm{e}^{i\,M\,\hat{\sigma}_z \theta_p /\, 2}\hat{f}(s)
\end{equation}
we can rewrite the quasiclassical equations (\ref{eq:QCAE})
in the form (\ref{eq:QCE0}) with the gap operator
\begin{equation}\label{eq:QCE0-DM}
    \hat{\Delta}_b(s) = \hat{\sigma}_x\,\mathrm{Re}D_M(s) %
                       -\hat{\sigma}_y\,\mathrm{Im}D_M(s)\,.
\end{equation}
The symmetry properties of both the gap operator $\hat{\Delta}_b(s)$
and the equation (\ref{eq:QCE0}) depend on the vorticity $M$:
\begin{equation}\label{eq:QCE2-SYMGM}
\hat{\Delta}_b(-s)= \left\{
        {\;\;\;\hat{\Delta}_b^*(s)\,,\quad \mathrm{for\; even}\,\, M}
        \atop%
        {-\hat{\Delta}_b^*(s)\,,\quad \mathrm{for\; odd}\,\, M}
           \right.\,.
\end{equation}
This fact allows to obtain the following condition:
\begin{equation}\label{eq:QCE2-SYMM}
\hat{f}(-s)= \left\{
        {C\, \hat{\sigma}_x \hat{f}(s)\,,\quad \mathrm{for\; even}\,\, M}
        \atop%
        {C\, \hat{\sigma}_y \hat{f}(s)\,,\quad \mathrm{for\; odd}\,\, M}
           \right.\,,
\end{equation}
which generalizes the condition (\ref{eq:QCE0-SYM}) for a multiquantum vortex.
Here $C$ is an arbitrary constant.
Using the stationary phase method
we can write the boundary condition
for wave functions $\hat{f}(s)$
at the surface of the insulating cylinder
in the form:
\begin{equation}\label{eq:WFBC2M}
    \mathrm{e}^{ i\hat{\varphi}_M}\hat f(s_0)=
    \mathrm{e}^{-i\hat{\varphi}_M}\hat f(-s_0)\, ,
\end{equation}
where
$$
\hat{\varphi}_M = k_\perp s_0 +
    (2\mu + M \hat\sigma_z)\beta_0 - 3\pi/4\, .
$$
Taking into account the Eq.(\ref{eq:QCE2-SYMM})
the boundary condition (\ref{eq:WFBC2M}) can be written
for electron $f_u$ and hole $f_v$
components of the wave function $\hat{f}$:
\begin{equation}\label{eq:WFBC3M}
f_v(s_0)= \pm e^{i M \alpha_0} f_u(s_0).
\end{equation}
For $\vert\, b\, \vert > R$ we can put here $s_0 = 0$ and
$\alpha_0 = -\pi/2$.
The choice of the sign in (\ref{eq:WFBC3M}) depends on the number of the spectral branch.
The typical plots of quasiparticle spectra obtained
from numerical solution of
the eigenvalue problem (\ref{eq:QCE0}), (\ref{eq:QCE0-DM})
with the boundary condition (\ref{eq:WFBC3M})
for vortices with winding numbers $M=2,3$
are shown in Fig.~4.
Similarly to the case of a singly quantized vortex the
small $b$ part of the spectrum is formed by the spectral branches induced by the normal scattering at the defect.
These branches transform into the
standard anomalous ones with the increase in the $\vert\, b\, \vert$ value.
With the increase in the
cylinder radius all the spectral branches appear to be expelled from the Fermi level.

\section{Summary}\label{sum}
To sum up, we described a transformation of the subgap spectral branches
of quasiparticle excitations in vortices pinned by columnar
defects of finite radii.
We find that the normal scattering at the defect surface results in the appearance of
additional spectral branches which transform into the CdGM one with an increase in the impact parameter of quasiparticle trajectories.
The increase in the defect radius is accompanied by the increase in the minigap in the spectrum which can be observed, e.g.,
in the STM measurements.
One can expect that such changes in the spectrum behavior should affect strongly the dynamic mobility of vortices in the presence
of ac transport current (see, e.g., \cite{Kopnin-TNSC} for review).


\section*{ACKNOWLEDGEMENTS}\label{Acknow}
We are thankful to N.\ B.\ Kopnin, A.\ I.\ Buzdin, V.\ M.\ Vinokur, and G.\ Karapetrov for stimulating discussions.
 This work was supported, in part,  by the
Russian Foundation for Basic Research  and by the ``Dynasty'' Foundation (A.S.M.).

\end{document}